\documentclass[11pt,twoside]{article}
\usepackage{Blair_asp2004}
\usepackage{psfig}
\usepackage{epsf}
\usepackage{graphics}
\usepackage{lscape}
\def\edcomment#1{\iffalse\marginpar{\raggedright\sl#1\/}\else\relax\fi}
\reversemarginpar

\begin{document}
\footnotesize
\noindent
To appear in ``1604-2004: Supernovae as Cosmological 
Lighthouses," a conference held June 2004 in Padua, Italy; 
to be published as an ASP Conference Proceedings.\\

\normalsize
\title{Kepler's Supernova Remnant:  The View at 400 Years}
\author{W. P. Blair}
\affil{Department of Physics \& Astronomy, The Johns Hopkins University, 
3400 N Charles St., Baltimore, MD, 21218 USA}

\begin{abstract}
October 2004 marks the 400th anniversary of the sighting of SN 1604,
now marked by the presence of an expanding nebulosity known as
Kepler's supernova remnant. Of the small number of remnants of
historical supernovae, Kepler's remnant remains the most enigmatic.
The supernova type, and hence the type of star that exploded, is 
still a matter of debate, and even the distance to the remnant is 
uncertain 
by more than a factor of two.  As new and improved multiwavength
observations become available, and as the time baseline of observations
gets longer, Kepler's supernova remnant is slowly revealing its
secrets.  I review recent and current observations of Kepler's
supernova remnant and what they indicate about this intriguing object.

\end{abstract}
%\footnote{This is a footnote example. Just insert in text.}
\thispagestyle{plain}

\vspace{-0.4in}

\section{Introduction}

Four hundred years ago, when Johannes Kepler and others observed the 
``new star'' of 1604, those observing the event had no concept of what it was
that they were observing.  Today we know that supernovae are exploding stars
and that they even come in different varieties.  The type Ia SNe, which 
have become so important for cosmology because they are standard candles, 
arise from the incineration of white dwarf stars.  The other main class,
core collapse SNe, come from more massive stars and produce the sub-classes
called type Ib, type Ic, and type II.  Of the six historical SNe in our Galaxy
that have occurred over the last millennium (including Cas A for which 
the SN itself apparently escaped detection), only Kepler's SN has remained 
uncertain as to the type of the star that exploded.

Today of course, what we observe is the expanding young supernova remnant 
(SNR) that resulted from the explosion.  By studying this SNR across the
electromagnetic spectrum, modern astronomers are still trying to discern
a clear picture of the precursor star of this event. In doing so, they are
hindered by the lack of an accurate distance to the object, which makes
the derivation of even basic properties like the diameter or mean 
expansion velocity uncertain.  In this paper, I will review the current 
observational status of this enigmatic object and point out some of the 
apparent inconsistencies in existing interpretations.

\section{Observational Parameters}

Kepler's SNR is located at galactic coordinates $l=4.5^{\circ}$ and
$b=6.8^{\circ}$ (i.e., nearly directly toward the galactic center from 
the sun but significantly out of the plane). The full extent of the remnant 
is most visible in radio and X-ray regimes, where a circle of diameter
200\arcsec\ encompasses the entire shell except for two ``ears'' of
emission on the east and west sides (see Figure 1).

Unfortunately, the distance to Kepler's SNR is only poorly constrained 
by observations to date.  Most recent literature cites the study by 
Reynoso \& Goss (1999) and adopts a distance near 5 kpc.  However, a careful 
reading of this paper shows that many authors misquote or misunderstand
Reynoso \& Goss's result. These authors use the H~I kinematics with a
galactic rotation model to place a rather inaccurate ``lower limit''
of 4.8 $\pm$ 1.4 kpc on the distance, and independently place an
``upper limit'' of 6.4 kpc based on the proposed association of the
SNR with an H~I cloud.  Kinematic distances are inherently uncertain
along the line toward the galactic center.  Table 1 shows how some basic
parameters for the SNR depend critically on the assumed distance.
In particular, note that the larger distances imply a very large distance
off the galactic plane and, when combined with the
observed current shock velocity, apparently require a very significant
deceleration, which is not consistent with the absence of a well-defined
reverse shock in the X-ray data.

\begin{table}[!ht]
\begin{center}  
\caption{Kepler's SNR: The Affect of the Distance Uncertainty}
\smallskip
{\small
\begin{tabular}{cccc}
\tableline
\noalign{\smallskip}
Parameter & D=3.0 kpc & D=4.5 kpc & D=6.0 kpc \\
\noalign{\smallskip}
\tableline
\noalign{\smallskip}
Z Distance (pc)    & 355  & 533  & 710 \\
Radius (pc)        & 1.45 & 2.18 & 2.90 \\
$<V_{exp}> ~\rm (km~s^{-1}$)  & 3540 & 5310 & 7080 \\
\noalign{\smallskip}
\tableline
\end{tabular}
}
\end{center}
\end{table}

In Figure 1, I show 6 cm VLA radio (DeLaney et al. 2002) and 
0.2 - 10 keV {\it Chandra} X-ray observations (Hwang et al. 2000) of 
Kepler's SNR.  The overall similarity is striking, showing a
roughly spherical thick shell of emission brightest in the north.
The apparent band of emission cutting across the middle from 
NW to SE is largely an illusion, caused by projection effects
from material on the front and back sides of the shell (see 
optical section below).

% Figure 1
\begin{figure}[!ht]
%\epscale{0.9}
\plottwo{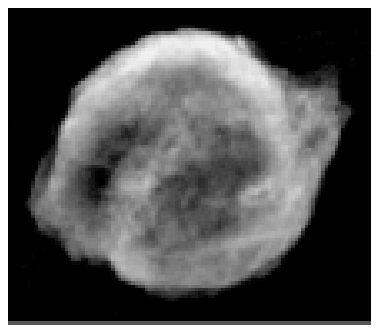}{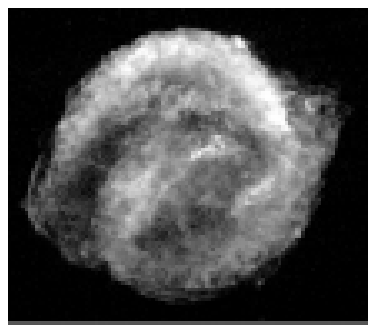}
\caption{\label{fig:fig1} 
(left) VLA 6 cm radio map of Kepler's SNR, from DeLaney et al. (2002). 
(right) {\it Chandra} 0.2 - 10 keV X-ray data from Hwang et al. (2000).
}
\end{figure}

However, this apparent similarity may be deceiving.  Flat and steep
radio spectrum deconvolutions look quite different from the total 
intensity maps, and soft (0.3 - 1.4 keV) and hard (4 - 6 keV) X-ray 
bands, either from {\it Chandra} or {\it XMM-Newton} (Cassam-Chena\"{i} et al. 2004)
also show different structures. (In particular, the harder X-rays
form a distinct outer rim likely associated with the primary shock wave.)
Also, when one looks at the dynamics of the SNR, discrepancies are seen.  
Both radio and X-ray observations extend
over a long enough baseline that expansion of the SNR has been measured.
Hughes (1999) finds an X-ray expansion rate of R $\propto ~ t^{0.93}$,
which is nearly free expansion.  This is almost twice that found in the
radio (R $\propto ~ t^{0.50}$, Dickel et al. 1988; DeLaney et al. 2002).
The reason for this discrepancy is not understood.

Optical observations (Blair, Long, \& Vancura 1991) are brightest in the 
NW, with the northern cap and isolated central patches of emission also 
visible (see Figure 2).  The optical data indicate 
substantial and variable foreground extinction, with E(B-V) = 1.0
$\pm$ 0.2, consistent with and X-ray determined N(H) = 5.0 $\times ~
10^{21} ~ \rm cm^{-2}$. Bandiera \& van den Bergh (1991) performed a careful
study of the space velocity of the object, finding a value of 278
$\rm km ~ s^{-1}$ toward the NW, which is away from the galactic plane.
Assuming this motion is due to the precursor star, this would account 
for the large angular distance off the galactic plane and 
the observed morphology in all bands, showing the brightest emission
in the N and NW.

% Figure 2  
\begin{figure}[!ht]
%\epsscale{0.8}
\plotone{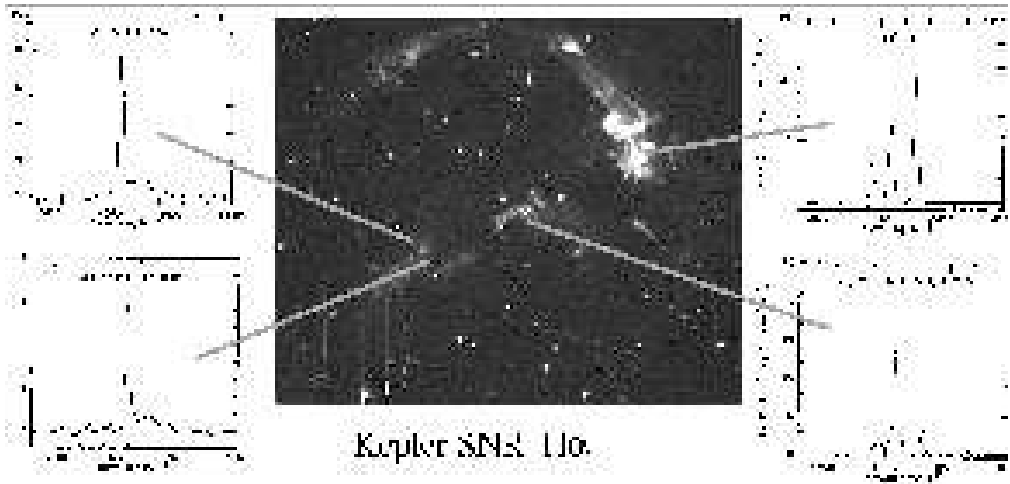}
\caption{\label{fig:fig2}
Ground-based H$\alpha$ image of Kepler's SNR (from Blair et al. 1991) is
shown.  A continuum image has been subtracted although some stellar
residuals remain.  Insets show H$\alpha$ + [N~II] region 
optical spectra of the regions indicated.  Note the broad and narrow
components on the H$\alpha$ line at some positions. Blue-shifted broad 
components indicate approaching (near side of shell) emission while
red-shifted material is from the back of the shell.
}
\end{figure}

The optical emission comes from two components: radiative shocks into
dense, knotty structures (presumably circumstellar mass loss), and
smoother filamentary emission visible only in H$\alpha$ from so-called
nonradiative shocks (e.g. Blair et al. 1991; Sollerman et al. 2003).
The radiative knots show enhanced [N~II]/H$\alpha$, indicative 
of probable enrichment from the precursor.
Interpretation of the broad line components in the nonradiative shocks
(e.g. Chevalier, Kirshner, \& Raymond 1980) indicate a current shock 
velocity of about 1750 $\pm$ 250 $\rm km ~ s^{-1}$. 

Relatively little information is available in the infrared.  The SNR
is small enough in angular size that {\it IRAS} data are not useful. 
However, Douvion et al. (2001) observed Kepler's SNR with thr
{\it Infrared space Observatory's} ISOCAM
(see Figure 3).  The morphology of the $\sim$12 $\mu$m emission is
very similar to the optical, and 6 - 16 $\mu$m spectra are well-fitted by
shock-heated dust models with T$_{dust}$ = 95 - 145 K, n$_{e}$ of several
thousand (similar to optical [S~II] densities), and T$_e$ of several
hundred thousand K.  This makes it very likely that the {\it ISO} emission
arises primarily from dust heated directly by the primary shock front.
SCUBA sub-mm observations at 450 $\mu$M and 850 $\mu$m by Morgan et al.
(2003) were interpretted as indicating a large mass ($\sim1 ~ M_{\odot}$
of cold (T = 17 K) dust
in the remnant, but this result has been called into question (Dwek 2004).
Upcoming {\it Spitzer Space Telescope} observations may resolve this issue.

% Figure 3  
\begin{figure}[!ht]
%\epsscale{0.6}
\plotone{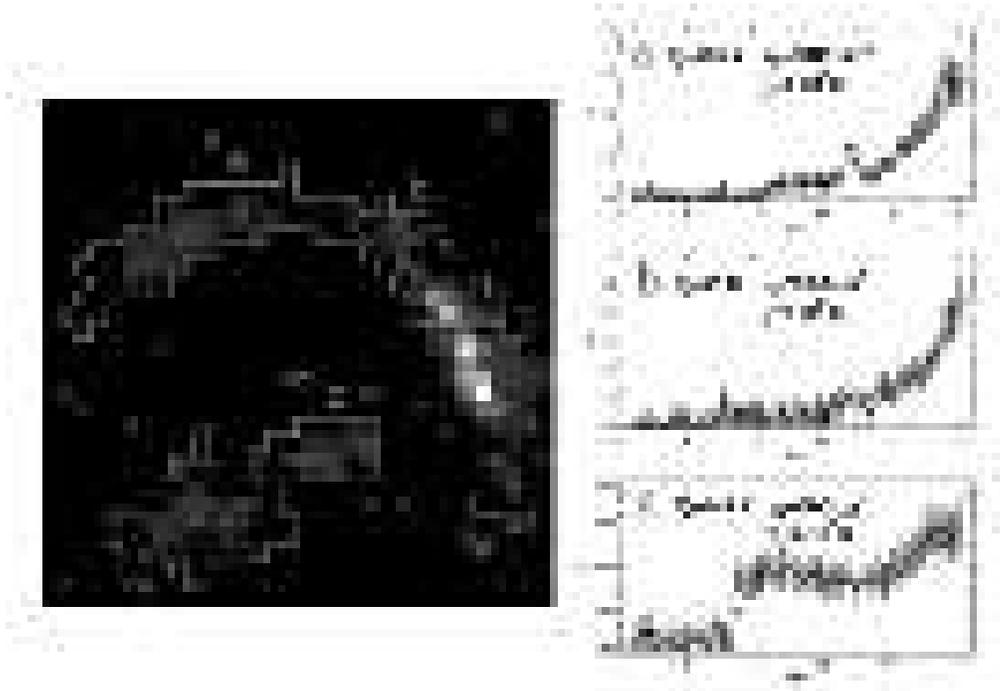}
\caption{\label{fig:fig3}
ISOCAM 14 - 16 $\mu$m image of Kepler's SNR (from Douvion et al.
2001). Boxes mark regions for extracted spectra, which show warm (95 -145 K)
thermal dust emission. The emission in boxes a-e accounted for 92\% of
the detected emission.  Note the similarity to the distribution of optical
emission in Figure 2.
}
\end{figure}

\section{Whither the SN Type?}

The literature is extremely confusing on the issue of the SN type. 
Baade (1943) reconstructed the historical light curve and claimed a
type Ia designation, a result that is still quoted in many recent papers.
However, Doggett \& Branch (1985) showed consistency with a type II-L
light curve, and Schaefer (1996) has also called the historical curve
and type Ia designation into question.

A decade ago, the preponderance of evidence seemed to point toward
a core-collapse event, and much of this evidence is still relevant.
A large distance off the galactic plane might be suggestive of a white
dwarf precursor, but high space motion away from the plane and N-rich 
circumstellar material points to a massive runaway star from an earlier
SN as the precursor for Kepler's SNR.  Borkowski et al. (1992, 1994)
developed a massive star model that was consistent with observations 
available at that time.  However, X-ray analyses in particular, from 
Exosat (Smith et al. 1989; Decourchelle \& Ballet 1994; Rothenflug 
et al. 1994), ASCA (Kinugasa \& Tsunemi 1999), 
{\it Chandra} (Hwang et al. 2000), and now {\it XMM-Newton} (Cassam-Chena\"{i} et al.
2004) have alternately claimed better fits to type Ia or core collapse
models.

% Figure 4  
\begin{figure}[!ht]
%\epsscale{0.6}
\plotone{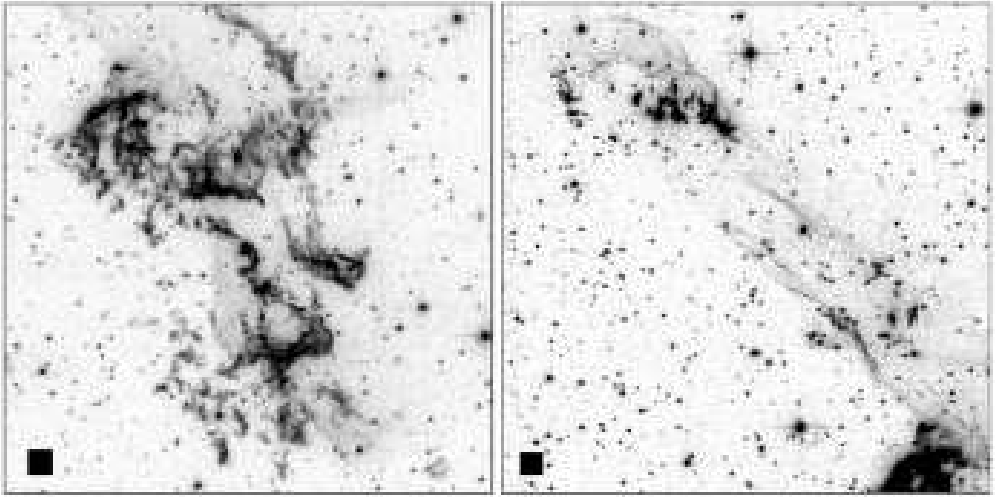}
\caption{\label{fig:fig4}
Two sub-fields in the northwestern portion of Kepler's SNR, as
seen with the {\it HST} Advanced Camera for Surveys, using filter F680N
(H$\alpha$+[N~II]). Bright knots (left panel) show the encounter of 
the blast wave with dense circumstellar knots.  The fainter wispy 
filaments, especially visible at right, are
nonradiative shocks delineating the actual shock front position.
The black square at lower left in each panel is 2\arcsec.
Comparison of the motion of the wispy filaments to the ground-based
data in Figure 2, with assumptions about the shock velocity, will
yield a much better distance estimate in the near future. (Figure
courtesy R. Sankrit, JHU.)
}
\end{figure}

The modelling of {\it XMM-Newton} data by Cassam-Chena\"{i} et al. 
(2004) is the most careful and accurate to date.  Although these authors
do not claim to have determined the SN type, their determination of 
Si and Fe abundances similar to type Ia models 
and their lack of detection of overabundances of O, Ne, Ar, S as
seen in core collapse objects such as Cas A (Hughes et al. 2000), 
is a strong indicator
of a type Ia event.  I also note that, even with its exquisite sensitivity 
and resolving power, the {\it Chandra} observation has failed to detect any
hint of a stellar remnant (quite in contrast to the situation with Cas A!).
While the issue may not be closed, the pendulum has swung back toward
a white dwarf precursor star.
This is not an entirely comfortable situation.  If the type Ia 
designation is correct, then the closest example of our cosmological
standard candle has some very peculiar properties!

At least one significant near term advance is in the offing. Recent
optical {\it HST} ACS images have been obtained (see Figure 4) that not
only show the bright optical filaments in exquisite detail, but will
permit the proper motion of key filaments to be measured. With a
refined estimate of the shock velocity from X-ray and optical data,
it should soon be possible to measure the distance to Kepler's SNR
with relative precision. 

Johannes Kepler was an intriguing personality (Ferguson 2002).  He
came from extremely humble beginnings, he possessed a fierce and
staunch religious faith that was entwined with his world view but 
at odds with his contemporary culture, and he was a visionary 
scientist. While understanding the new star of 1604 was a sidelight
for him, I think he would be secretly pleased to know that his name
has been attached to this equally intriguing supernova remnant.

\acknowledgments{I thank the conference organizers for inviting me to present this
review and for partial support.  This work was also supported by
STScI grant GO-09731A to the Johns Hopkins University.}

\end{document}